\documentclass[twocolumn,showpacs,preprintnumbers,amsmath,amssymb]{revtex4-1}

\usepackage{graphicx}
\usepackage{dcolumn}
\usepackage{bm}

\begin{document}

\title{Auger mediated quantum sticking of positrons to surfaces: 
Evidence for single step transition from a scattering state to a surface 
image potential bound state}

\author{
S. Mukherjee$^1$, M. P. Nadesalingam$^1$, 
P. Guagliardo$^2$, A.D. Sergeant$^2$ , B. Barbiellini$^3$, 
J.F. Williams$^2$, N.G. Fazleev$^1,^4$, A.H. Weiss$^1$}

\affiliation{
$^1$ Department of Physics, University of Texas at Arlington, 
Arlington, TX, USA,76019\\ 
$^2$ ARC Centre of Excellence for Antimatter-Matter Studies,  School of Physics, The University of Western Australia, 
Crawley,WA 6009 Australia\\
$^3$ Department of Physics, 
Northeastern University, Boston, 
Massachusetts 02115, USA\\
$^4$Department of Physics, Kazan State University, Kazan, 420008, Russian Federation}


\begin{abstract} 
We present the observation of an efficient mechanism
for positron sticking to surfaces termed here Auger mediated quantum sticking.  
In this process the energy associated with the positrons transition from an unbound
scattering state to a bound image potential state is coupled to a valence electron 
which can then have sufficient energy to leave the surface. Compelling evidence for
this mechanism is found in a narrow secondary electron peak observed at incident positron 
kinetic energies well below the electron work function.

\end{abstract}

\pacs{78.70.Bj, 71.60.+z, 68.47.De, 03.65.Sq}

\maketitle


Recently positrons \cite{1} have been shown to be very effective in probing surfaces
and reduced dimensional systems such as nano-particles which possess high surface-to-
volume ratios.  
If positrons become trapped in image potential surface states before annihilation, 
they can provide a means of selectively sampling the top most layer of a material or nanostructure 
due to the fact that such states typically extend about one atomic layer
below the surface.  Subsequent annihilation of surface trapped positrons with core or valence electrons results in signals 
(e.g. annihilation induced Auger electrons  \cite{2} 
or annihilation gamma rays \cite{3}) containing crucial information about the composition of 
the outermost regions of nano-materials.

In this letter, we present experimental evidence for an efficient
mechanism for depositing positrons directly 
into surface states through a single quantum step. In
this process, the energy associated with the positron transition from an unbound scattering 
state to a bound surface state is coupled to a valence electron 
which may then have sufficient energy to leave the surface. 
Due to its similarity with the Auger transition in solids, this process has been
termed as Auger mediated quantum sticking (AMQS). The quantum nature of the AMQS follows 
from the fact that the de Broglie 
wavelength of a $1$ eV positron (about $12$ \AA) is an 
order of magnitude more than the width of the surface potential well (about 1 \AA ) \cite{7}.
Similar ideas have been suggested in theoretical models \cite{4,5,6}, however no
experimental evidence of this mechanism was available until now. 
The AMQS process schematized in Fig.~1 is related closely to 
the Auger de-excitation of atoms \cite{8}  
or molecules \cite{9} near surfaces, which has been studied 
for decades in various fields.

Here we provide direct experimental confirmation of the AMQS process through measurements
of electron energy 
spectra resulting from very low energy positron bombardment ($1.5-7$ eV).  
The strongest evidence
for the AMQS is found in a narrow electron peak
observed at incident positron kinetic energies well below the electron work function value. 
The present experiment also allowed us to determine the positron 
sticking probability as a function of incident particle energy 
and to obtain an independent measurement of 
the positron binding energy at the surface \cite{10}. 
The fact that this new mechanism has an efficiency exceeding $10$ \% at positron
energies $\sim1$ eV proves that it will be possible to use low energy positron beams 
to selectively probe the surfaces of fragile systems such as nano-particles and 
bio-materials and to obtain Auger signals 
which are completely free of secondary electron background.

\begin{figure} 
\begin{center} 
\includegraphics[width=\hsize,width=8.cm]{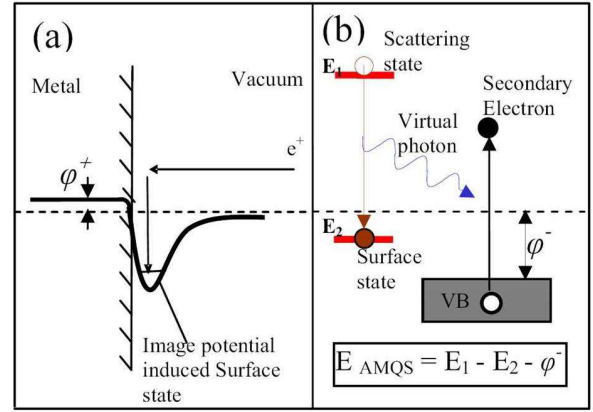} 
\end{center} 
\caption{ (Color online) Schematics of AMQS process:
(a) the slow positrons in real space is directly trapped at
the surface state resulting in the release of an electron carrying away the residual energy 
(b) the energy band diagram showing the AMQS (VB = valence band )}
\label{fig1} 
\end{figure}

 
The experiments were carried out using Time of Flight -Positron Annihilation 
induced Auger Electron Spectrometer (TOF-PAES) \cite{11} which uses a magnetic bottle analyzer\cite{12}. 
Positrons from a Na-22 source are guided to the sample 
using an \textbf{E}$\times$\textbf{B} field. 
The emitted electrons are detected by a 
multi channel plate (MCP).
The energy of the emitted electron was calculated from the electron time of flight (TOF) which was determined from the time difference between the electron and annihilation gamma ray detections (see Ref. \cite{11}). 
An Au sample (a $99.985$ \% pure polycrystalline foil, $0.025$ mm thickness) 
was sputter cleaned every $12$ hours while a Cu(100) sample 
(a $99.9$ \% pure single crystal $10$ mm diameter $\times$ 
$1$ mm thickness) 
was sputter cleaned followed by annealing at $740$ 
$^o$C every 12 hours. 
The incident beam profile at $0$ V sample bias was fitted with a Gaussian of $0.4$ eV FWHM 
and maximum at $0.65$ eV. $99$ \% of the positrons have energy 
less than  $1$ eV and this is referred to as the beam energy. The incident positron beam energy was increased by negatively biasing the sample.


%
%

The primary evidence for the AMQS process in Cu is shown in
Fig.~2(a), where
the normalized energy spectrum taken at different 
positron beam energy $E$ ($1.5$-$7$ eV) is plotted. 
In each spectrum the large peak at low energies ($< 10$ eV) corresponds 
to electrons that are emitted as a result of positron impact at 
the sample surface.  
The much smaller peak at about $60$ eV is the 
PAES \cite{2}.

\begin{figure} 
\begin{center} 
\includegraphics[width=\hsize,width=8.cm]{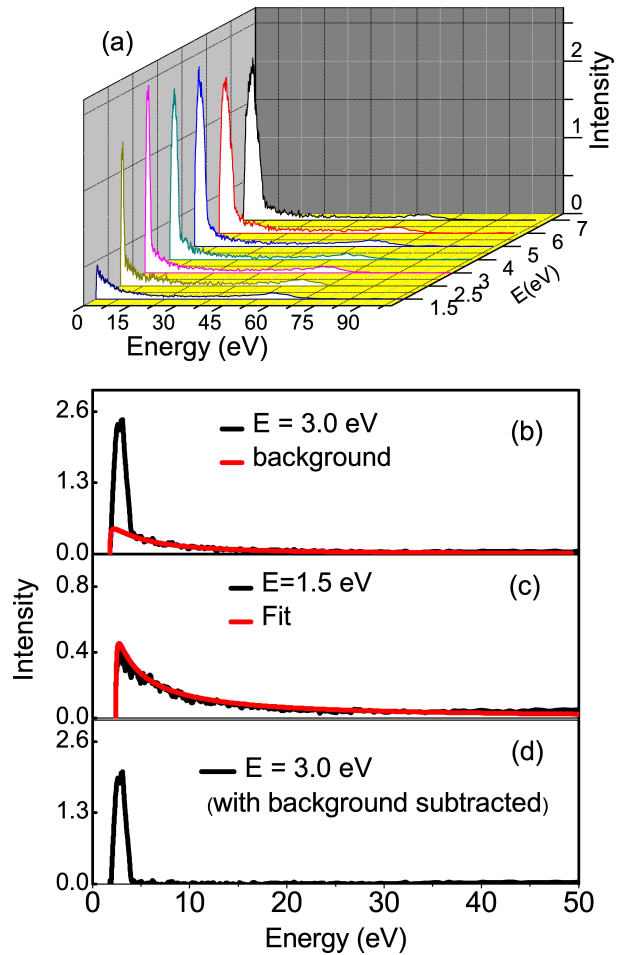} 
\end{center} 
\caption{(Color online)
Electron energy spectrum of emitted from Cu(100) resulting 
from low energy positron bombardment:(a)
The spectrum were taken for a series of positron energy ranging from 
$1.5$ eV to $7$ eV. 
All the data have been normalized to Auger peak. 
(b) Electron energy spectrum taken with a beam energy of $3$ eV;
(c) Energy spectrum with beam energy $1.5$ eV; (d) same as (b) 
but with background, as estimated from (c), subtracted.
} 
\label{fig2} 
\end{figure}
%
%

AMQS can be distinguished from another process 
in which the final state of the incident positron is the bulk state.
In the latter
process, the maximum kinetic energy of the outgoing electron
(as measured outside the sample surface) 
is given by 
\begin{equation}
E_{\mbox{max}}=E - \phi^- + \phi^+,
\end{equation}
where $E$ is the incident positron energy (measured from the vacuum level) and$\phi^-$ = 4.65 (4.8) eV is the electron work function  in Cu(Au) and $\phi^+$ = -0.02 (+0.9) eV is the positron work function [13-16]. Both work functions are measured from the vacuum level with positive sign below the vacuum level.
The electrons can escape the sample if $E_{\mbox{max}} > 0$ eV which implies  
that $E$ should be greater than $\phi^- - \phi^+$. 
Hence, for incident positron kinetic energies of 
less than $4.7 (3.9)$ eV there should be no secondary electron emission
according to this mechanism.  
In the case of AMQS, the positron excites an electron-hole pair 
while dropping to the surface state.
The energy to dissipate is the initial positron kinetic energy 
plus the positron binding energy to
the surface, thus we have 
\begin{equation}
E_{\mbox{max}}=E +E_{ss}- \phi^-,
\end{equation}                                                                      
where $E_{ss}$ is the surface binding energy of the positron measured from the vacuum level (with positive sign below the vacuum). 
Annihilation induced processes 
including Auger transitions and $\gamma$-ray emission 
can lead to the emission of electrons with energies as high as the Auger 
transition energy and $511$ keV respectively.  However, such processes 
would lead to the formation of broad electron peaks. 
In our experiments, we found a narrow electron peak even when the incident 
kinetic energies of the positrons were less than $3$ eV. 
This can be explained from Eq.~2
by considering the process in which the electron excited from the Fermi sea 
escapes from the surface if the positron incident energy is greater than a certain threshold 
of about $2$ eV since $E_{ss}$ is of the order of $3$ eV in most metals
\cite{15}.

%
%

Fig. 2(b) shows a typical electron spectrum of Cu.  The large peak centered 
at $3$ eV corresponds to AMQS induced electrons. Fig. 2(c) shows the spectrum 
similar to Fig. 2(b) except the beam energy is below the threshold for electron emission, thus,  
one can notice the absence of the low energy (AMQS) peak:
the quantum sticking is still 
taking place, but electron emission outside the sample is energetically prohibited. 
The same broad background between 
$5 - 30$ eV can be seen in both Figs 2(b) and 2(c). This 
feature is presumably the low energy electron tail associated with the PAES peak at  $60$ eV.  
Fig. 2(d) shows the same spectrum of Fig. 2(b) but with the PAES background subtracted.

%
%

As a test to determine whether the low energy peak (AMQS peak) is due to 
a secondary effect of PAES or not, we have heated up the sample 
to $700$ $^o$C and  performed measurements with 
incident positron beam energy above (Fig. 3(a)) and below (Fig. 3(b)) 
the threshold given by Eq.~2 respectively. 
At $700$ $^o$C, 
the positron trapped in the surface state is desorbed  as positronium \cite{15}.
This prevents  annihilation of the positron with core electrons and eliminates the PAES peak. 
Therefore, the presence of the AMQS peak at high temperature
in Fig. 3(a) proves that this feature is not 
associated with the PAES process. 
Finally, the PAES peak always disappears 
at high temperature (as shown in Figs. 3(a) and 3(b)) 
while the AMQS peak disappears 
only when the beam energy is below threshold as given by Eq.~2.

\begin{figure} 
\begin{center} 
\includegraphics[width=\hsize,width=8.cm]{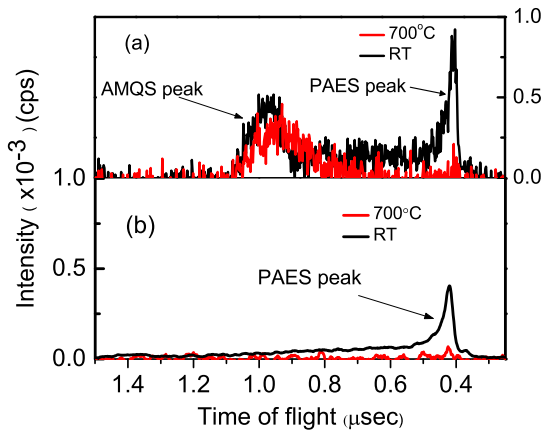} 
\end{center} 
\caption{(Color online)
Comparison of electron time of flight spectrum 
resulting from positron bombardment from hot and cold Cu(100) surfaces:
(a) Cu at  $700$ $^o$C and room temperature with a beam energy of $3$
eV (b) same as (a) but with a beam energy of $1.5$ eV. In (a)  when the sample is at room temperature the 
incident positrons stick to the surface via AMQS followed by annihilation with the core electrons resulting in PAES 
signal. When the positrons encounter the hot surface, they still undergo AMQS but are desorbed as positronium 
before they can annihilate with core electrons\cite{10} (note the inverted scale on TOF axis).
} 
\label{fig3} 
\end{figure}

%
%
The AMQS peak integrated intensity with  the background subtracted
is plotted in Fig. 4 (a) and (b) as a function of the
incident positron energy and it is used to estimate 
the surface state binding energy $E_{ss}$. Previous measurements of 
$E_{ss}$  needed the monitoring of the fraction $f(T)$ of 
incident positrons which forms positronium  as a function 
of sample temperature $T$ \cite{10}. 
Here, we fit the low energy part ($< 5$ eV) of the
AMQS peak integrated intensity
with a linear function of the positron incident energy. 
The intercept of the straight line
in conjunction with Eq.~2 is used to determine $E_{ss}$ for
Cu ($2.79  \pm 0.2$ eV)
and for Au ($2.87 \pm 0.22$eV).
This simple determination of $E_{ss}$ agrees well with 
values reported in literature \cite{16}.

%
%

The AMQS peak integrated intensity was also used to estimate the positron sticking probability $S(E)$. 
It has been assumed that the transition of the positron to the surface state is always associated 
to an electron-hole pair excitation. 
Hence, the sticking probability can be written as 
\begin{equation}
S(E)=\frac{N_{AMQS}/N_{e+inc}}{P(E)r(E)},
\end{equation} 
where $N_{AMQS}$ 
is the integrated 
intensity of the AMQS peak, 
$N_{e+inc}$ is the number of incident positrons,  
$P(E)$ is the escape probability for the excited electrons  \cite{18} 
and  $r(E)$ is that fraction of the excited electrons which have enough energy to escape 
by overcoming the workfunction $\phi^-$. The ratio $r(E)$ is therefore given by 
\begin{equation}
r(E) =\frac{ \! \int_{E_a}^{E_{F}}  g(E)\,dE}{ \! \int_{E_b}^{E_{F}}  g(E)\,dE}
\label{eq:fine},
\end{equation}
where $E_F$ is the Fermi energy of the metal, 
$E_{a}=E_F-(E+E_{ss}-\phi^-)$,
$E_{b}=E_F-(E+E_{ss})$
and $g(E)$ is the density of states (DOS) \cite{19}. 
The number of incident positrons $N_{e+inc}$
was estimated using
\begin{equation}
N_{e+inc}=N_{SS}+N_{Ps}+N_{ref},
\end{equation}
where $N_{ss}$ is the number of positrons trapped in the surface state,   
$N_{Ps}$ is the number of positrons that form positronium and   
$N_{ref}$ is the number of positrons that are reflected from the surface.
$N_{Ps}$  and  $N_{ss}$ are related by
$N_{Ps}=f(N_{ss}+N_{ref})/(1-f)$ where $f$ is the fraction of incident 
positrons that form positronium ($\sim 0.5$, determined as  in Ref. \cite{10}),  while $N_{PAES}=C N_{ss}$  
where $N_{PAES}$ is the integrated intensity of PAES peak 
and  $C$ is the probability that a 
positron trapped in the surface state will annihilate 
with a core electrons($4.6$\%) \cite{14}. 
Taking $0.2$ as the upper limit of $R=N_{ref}/N_{e+inc}$ \cite{20}, 
the total number of incident positron can be written as   
\begin{equation}
N_{e+inc}=\frac{N_{PAES}}{0.8(1-f-R)CT_{PAES}},
\end{equation}
where  $T_{PAES}=0.45$ $(0.58)$  
is the fraction of Auger electrons for Cu (Au) that are transmitted to and detected 
by our analyzer. The sticking probability $S(E)$ estimated in this way is plotted in Fig 4(c) and its trend agrees well with the calculations by Walker et al. \cite{6}. This model predicts that $S(E)$ will vanish as $E\rightarrow0$. However, this prediction cannot be resolved by the present experiment. Here we only focus on high positron surface sticking rates for non zero positron energies.

\begin{figure} 
\begin{center} 
\includegraphics[width=\hsize,width=8.cm]{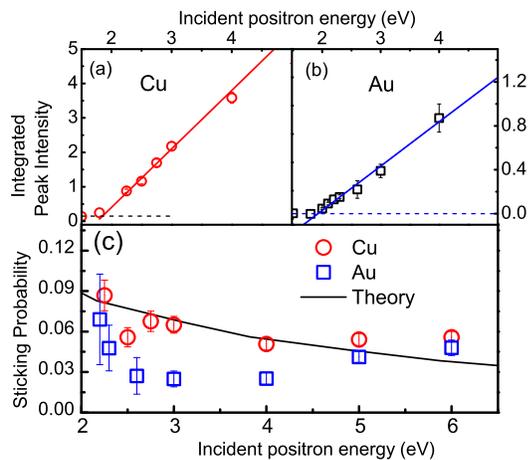} 
\end{center} 
\caption{(Color online)
AMQS peak integrated intensity for (a) Cu and (b) Au as a function of incident positron energy. 
The data for positron energy less then $5$ eV has been fitted with a straight line and 
the energy intercept has been used to determine the binding energy of the surface state \cite{17}. 
The dashed line corresponds to the average background when the AMQS process in energetically forbidden; 
(c) Sticking probability $S(E)$ calculated from Eq.~3. 
The theoretical calculation is from 
Ref.\cite{6} and corresponds to a typical electron 
density of $r_s=2$ au,
 $\phi^+=-0.16$ eV and 
 $E_{ss}=3$ eV.} 
\label{fig4} 
\end{figure}

We have reported experiments that provide strong
experimental evidence for a quantum sticking process in which an
incident low energy positron makes a direct transitions from an
unbound scattering state into an image potential surface state
resulting in emission of a secondary electron.  We have termed this
process Auger mediated quantum sticking (AMQS) because the energy lost as the positron makes a transition to the bound state is given to an outgoing electron.  Measurements of the incident beam energy at which the secondary peak first appears indicate a threshold almost $3$ eV lower than the value that would be expected if the positron were making a transition to a bulk state.  These measurements were used to obtain the first estimates of the surface state binding energy at room temperature.  The AMQS peak integrated intensity was used to estimate the sticking probability of positrons to surface and is found to be in qualitative agreement with the theory \cite{6}.  Our measurements provide the first
experimental demonstration of the trapping of positrons into the 
surface state with high efficiencies ($\sim10$\%) at incident positron
energies  below the threshold for collision induced secondary electron generation ($\sim{1.5}$ eV).
We have used this effect to obtain the first Auger spectra that are
completely free of background due to primary beam induced secondary
electrons. These measurements also demonstrate the possibility of greatly reducing the beam induced surface damage associated with Auger analysis by using incident positron energies below the threshold of chemical bond breaking \cite{21}.  The strong signal associated with the AMQS process suggests that measurements of the AMQS intensity as a function of energy can provide an independent way of testing models for inelastic scattering and quantum sticking of light particles \cite{22,23}.
        
We acknowledge useful discussions with A.P. Mills Jr., P.M. Platzman and R.M. Nieminen.
This work was supported in part by the NSF Grant No.$DMR-0907679$, Welch Foundation Grant No. Y-$1100$ and the Ministry of Education and Science of the Russian Federation Grant No. $2.1.1/2985$ and $2.1.1/3199$.  
B.B. was supported by Contract No. DE-FG02-07ER46352
from the Division of Materials Science and Engineering, 
Office of Science, U.S. Department of Energy
and benefited from the allocation of computer 
time at the NERSC and the Northeastern University's 
Advanced Scientific Computation Center.


\begin{thebibliography}{99}

\bibitem{1}
S. Eijt {\em et al.}, Nat. Mater. {\bf 5}, 23 (2006).
\bibitem{2}        
A.H. Weiss {\em et al.}, Phys. Rev. Lett. {\bf 61}, 2245 (1988).
\bibitem{3}
P. Asoka-Kumar {\em et al.}, Phys. Rev. Lett. {\bf 77}, 2097 (1996).
\bibitem{4}        
R. M. Nieminen in {\em Positron beam and their applications} 
edited by P. G. Coleman, 
World Scientific Singapore (2000).
\bibitem{5}
M.J. Puska and R.M. Nieminen, Rev. Mod. Phys. {\bf 66}, 841 (1994).
\bibitem{6}        
A. B. Walker {\em et al}.,
Phys. Rev. B {\bf 46}, 1687 (1992). 
\bibitem{7}        
A.P. Mills Jr, E.D. Shaw, M. Leventhal, P.M. Platzman, 
Phys. Rev. Lett. {\bf 66}, 735 (1991).
\bibitem{8}        
H. D. Hagstrum, 
Phys. Rev. {\bf 96}, 336 (1954).
\bibitem{9}        
B. Barbiellini and P.M. Platzman, 
New J. Phys. {\bf 8}, 20 (2006).
\bibitem{10}        
A. P. Mills Jr, Solid State Comm.  {\bf 31}, {\bf 623} (1979); K. G. Lynn, Phys. Rev. Lett. {\bf 43}, 391 (1979).
\bibitem{11}        
S. Xie, PhD Thesis, UT Arlington (2002). 
\bibitem{12}        
P. Kruit and F. H. Read, J. Phys. E {\bf 16}, 313 (1983).
\bibitem{13}        
A. P. Knights, P. G. Coleman,  
Surface Science {\bf 367}, 238 (1996); M. Farjam and H.  B. Shore, 
Phys. Rev. B {\bf 36}, 5089 (1987).
\bibitem{14}
N. Fazleev et al., Surface Science, {\bf 604}, 32 (2010).
\bibitem{15}  
A. P. Mills Jr, proceedings LXXXIII International School of Physics Enrico Fermi, ed. by W. Brandt, 
A. Dupasquier, Academic Press, New York (1982). 
\bibitem{16}        
M.J. Puska, R. M. Nieminen, Physica Scripta {\bf T4}, 79 (1983).
\bibitem{17}        
The error bars are predominantly due to the uncertainty in the intercept as measured in figures 4(a) and (b). 
\bibitem{18}
 $P(E)$  is from Ref. \cite{8} with $\alpha=0.7$ and $\beta=6$. 
\bibitem{19}        
We have assumed a constant DOS near $ E_{F}$.
\bibitem{20}        
J. A. Baker, M. Touat and P.G. Coleman, 
J. Phys. C: Solid State Phys. {\bf 21}, 4713 (1988).
\bibitem{21}        
A.H. Weiss {\em et al.}, Rad. Phys. and Chem. {\bf 76}  
285 (2007); A. H. Weiss in {\em the Proceedings of the International School of Physics 
Enrico Fermi, Physics of Many Positrons}, IOP Press to be published.
\bibitem{22}        
K. W. Goodman and V. E. Henrich, 
Phys. Rev. B {\bf 49}, 4827 (1994).
\bibitem{23}
A. P. Mills Jr. and P.M. Platzman in  {\em New direction in antimatter chemistry and physics}, 
edited by Clifford M. Surko and Franco A. Gianturco, 
Kluwer Academic Publishers (2001).

\end{thebibliography}
\end{document}